\documentclass[12pt]{article}

\pdfoutput=1

\usepackage{graphics}
\usepackage{epsfig}
\usepackage{amsfonts}
\usepackage{amssymb}
\usepackage{amsmath}
\usepackage{dsfont}
\usepackage{pifont}
\usepackage{bbm}
\usepackage{multirow}
\usepackage{latexsym}
\usepackage{verbatim}
\usepackage{mcite}
\usepackage{cite}
\usepackage{units}
\usepackage[all]{xy}
\usepackage{cancel}
\usepackage{slashed}
\usepackage{tikz}
\usetikzlibrary{snakes}
\newcommand{\be}{\begin{equation}}
\newcommand{\ee}{\end{equation}}
\newcommand{\ben}{\begin{displaymath}}
\newcommand{\een}{\end{displaymath}}
\newcommand{\bea}{\begin{eqnarray}}
\newcommand{\eea}{\end{eqnarray}}

\newcommand{\bean}{\begin{eqnarray*}}
\newcommand{\eean}{\end{eqnarray*}}

\DeclareMathAlphabet{\mathpzc}{OT1}{pzc}{m}{it}

\begin{document}
\pagestyle{plain}

%----------------------------------------------------------------------%
%  numbering sections, equations, footnotes, etc...
%----------------------------------------------------------------------%

\makeatletter \@addtoreset{equation}{section} \makeatother
\renewcommand{\thesection}{\arabic{section}}
\renewcommand{\theequation}{\thesection.\arabic{equation}}
\renewcommand{\thefootnote}{\arabic{footnote}}

%----------------------------------------------------------------------%
%  Resetting of counters
%----------------------------------------------------------------------%

\setcounter{page}{1} \setcounter{footnote}{0}

%----------------------------------------------------------------------%
%  title page
%----------------------------------------------------------------------%

\begin{titlepage}

\begin{flushright}
UUITP-27/16\\
\end{flushright}

\bigskip

\begin{center}

\vskip 0cm

{\LARGE \bf The fate of stringy AdS vacua and the WGC} \\[6mm]
%{\LARGE \bf On non-supersymmetric AdS vacua and their \\[2mm] universal instabilities} \\[6mm]

\vskip 0.5cm

{\bf Ulf Danielsson and\, Giuseppe Dibitetto \,  }\let\thefootnote\relax\footnote{{\tt ulf.danielsson@physics.uu.se, giuseppe.dibitetto@physics.uu.se}}\\

\vskip 25pt

{Institutionen f\"or fysik och astronomi, University of Uppsala, \\ Box 803, SE-751 08 Uppsala, Sweden \\[2mm]}

\vskip 0.8cm

\end{center}

\vskip 1cm

\begin{center}

{\bf ABSTRACT}\\[3ex]

\begin{minipage}{13cm}
\small

The authors of arXiv:1610.01533 have recently proposed a stronger version of the weak gravity conjecture (WGC), based on which they concluded that all those non-supersymmetric AdS vacua that can be
embedded within a constistent theory of quantum gravity necessarily develop instabilities. 
In this paper we further elaborate on this proposal by arguing that the aforementioned instabilities have a perturbative nature and arise from the crucial interplay between the closed and the open string 
sectors of the theory.

\end{minipage}

\end{center}

\vfill

\end{titlepage}

%%%%%%%%%%%%%%%%%%%%%%%%%%%%%%%%%%%%%%%%%%%%%%%%%%%%%%%%%
%%
%%               Contents
%%
%%%%%%%%%%%%%%%%%%%%%%%%%%%%%%%%%%%%%%%%%%%%%%%%%%%%%%%%%

\tableofcontents

\section{Introduction}
\label{sec:introduction}

An important goal for string theory as a consistent theory of quantum gravity, is to understand and control the mechanism of dynamical supersymmetry breaking in order to produce stringy 
embeddings of low energy effective models that describe phenomena at the particle physics scale. These comprize gravitational vacuum solutions as well as local models for particle phenomenology. 
Focusing on the former ones, the understanding of supersymmetry breaking turns out to be a crucial ingredient when you want to 
construct dS vacua with a solid stringy origin. The existence of such vacua could be very relevant for cosmological applications, while non-supersymmetric AdS vacua, apart from being
interesting per se, may have holographic relevance thanks to the AdS/CFT correspondence \cite{Maldacena:1997re}. 

If we for the moment neglect  the possible use of thermal fluctuations and focus on physics at zero temperature, a candidate mechanism for describing supersymmetry breaking in an AdS vacuum was proposed
in \cite{Argurio:2007qk}, where spacetime-filling anti-branes are created through motions in the underlying brane system resulting in an additional metastable vacuum.
From the viewpoint of the brane system, the above $\textrm{AdS}_{5}$ vacuum can be thought of as the near-horizon geometry of a thick extremal but non-BPS membrane obtained from a compactification of
a stringy brane intersection involving spacetime-filling anti-branes. 

Recently, in \cite{Ooguri:2016pdq}, a stronger version of the weak gravity conjecture (WGC) was proposed according to which such extremal but non-BPS membranes are unstable with respect to
nucleation of microscopic charged membranes. The immediate consequence is the instability of the corresponding AdS vacuum obtained by taking the membrane's near-horizon limit. 
Moreover, this instability would even happen instantaneously due to an infinite redshift effect affecting near-horizon observers. Additional consequences of this conjecture for the landscape have been
recently discussed in \cite{Freivogel:2016qwc}.

In the present work we want to take a further step, and extend the results to any AdS vacuum regarded as a flux vacuum in string- or M-theory, rather than constructed through the use of branes.
The key question is whether all AdS (flux) vacua may be thought of as arising by taking a near-horizon limit of a construction where the fluxes are supported by brane sources (see \emph{e.g.} \cite{Kounnas:2007dd}). 
The canonical example is $\textrm{AdS}_2 \times S^2$, which is obtained as the near-horizon limit of a Reissner-Nordstr{\"o}m (RN) black hole. If this is true in general, one can use information concerning 
the (in)stability of the brane picture to answer questions about stability of the corresponding AdS vacua. In fact, with or without the WGC, it is well known, and generally expected, that the spacetime of non-extremal, and even exactly extremal RN black holes, suffers from instabilities. In case of non-extremal black holes the inner horizon is destroyed, while the extremal black hole has its coincident horizon demolished. The conclusion is that the wormholes in the interior of the black holes are closed up and there is no escape from the singularity out into a new universe.

The AdS in the near-horizon limit of an extremal black hole makes crucial use of the wormholes.  This can be seen by following how the oscillatory motion of a freely falling observer goes back and forth through the horizon as shown in figure~\ref{fig:Warmhole}. From the perspective of the brane picture only half an oscillation is visible in any given universe. The next half a period takes the freely falling observer through a wormhole and out into a new universe through a white hole. This spacetime structure must remain intact if one wants to have a long lasting AdS.
This does not seem to be the case in general, since any field coupled to gravity at the horizon develops a perturbative instability \cite{Lucietti:2012xr,Aretakis:2011ha,Aretakis:2011hc}.  

In the case of a supersymmetric AdS vacuum, the stability of the corresponding brane picture suggests that the instability present in the general case is no longer there. This can be presumably understood using the very special particle spectrum of the supersymmetric theory that makes sure that any dangerous build up of energy near the horizon is cancelled.

Interestingly, there are hints that suggest that non-supersymmetric AdS vacua are completely stable as long as one restricts to the closed string sector. As shown in some explicit examples \cite{Danielsson:2016rmq}, the vacua turn out to be non-perturbatively stable and there is no tunneling to neighbouring vacua. At most one can have extremal domain walls (DW) separating different vauca, which, from the CFT point of view, correspond to renormalization group flows. From a calculational point of view these results are quite surprising. 

We believe that this is part of a larger picture where all AdS vacua obtained from $\mathcal{N}=8$ SUGR are in fact non-perturbatively stable. This echoes the conjectures by Tom Banks who claimed long ago that AdS should be absolutely stable in a sensible theory of SUGR \cite{Banks:2002nm}. So, what is then the status of $\mathcal{N}=8$ SUGR? According to \cite{Green:2007zzb} there is no limit that can be taken in string theory where all suprefluous states decouple and only $\mathcal{N}=8$ SUGR remains. There will always be some light states around that eventually bring you out of the theory. We believe that this could be exactly what the brane picture is all about. The AdS vacua obtained in the near-horizon limit can never forget their origin, and the instabilites generated through the WGC that force them to decay. The only way out would be if there is a different theory, without a brane picture, where the AdS is not obtained through such a limiting procedure. The candidate theory would be $\mathcal{N}=8$ SUGR that then would exist independent of any (nonexistent) limit of string theory.

The real question is whether such a theory can ever exist, which boils down to the well known question of the UV-finiteness of $\mathcal{N}=8$ SUGR. 
It is only, we claim, to the extent that $\mathcal{N}=8$ SUGR can stand on its own that holography can be true in an exact sense. Alternatively, all AdS vacua are obtained through a limit of a brane picture, and holography -- in the sense of mapping a CFT to AdS -- turns out to be imperfect for non-supersymmetric systems.

\section{An interplay between closed and open strings}

We start out by reviewing the argument in \cite{Ooguri:2016pdq}. First of all, the WGC in its original form states that, in a consistent theory of quantum gravity coupled to a $\textrm{U}(1)$ gauge field,
there needs to exist a fundamental particle whose mass $m$ and charge $q$ obey the bound
\be
\label{WGC}
\left(\frac{m}{M_{\rm Pl}}\right)^{2} \, \leq \, q^{2} \ ,
\ee
where $M_{\rm Pl}$ denotes the Planck mass. This allows one to get of rid of all those would-be remnants given by macroscopic extremal black holes, by simply offering them a viable decay channel
into light charged particles.

The sharpened version of the WGC proposed in \cite{Ooguri:2016pdq} further states that the inequality \eqref{WGC} can \emph{only} be saturated within a supersymmetric theory and if, furthermore, the 
particle in question is BPS. This conjecture was then used by the authors to conclude that any non-supersymmetric AdS vacuum emerging as near-horizon geometry of a non-BPS membrane will be unstable.
Such an instability is associated with the possibility of nucleating microscopic charged membranes obeying the bound in \eqref{WGC} in a strict sense.

Note that this channel of instability gives the membrane a finite lifetime when viewed from afar. When we zoom in on the horizon, in order to adapt the point of view of an observer in AdS, such a process will occur instantaneously
due to an infinite redshift effect. \footnote{It is argued in \cite{Ooguri:2016pdq} that even a finite lifetime in AdS will lead to an immediate destruction of the CFT-dual.} In what follows we will argue that the membrane approach to AdS, which allows one to draw this rather universal conclusion, is completely general and
corresponds to something physical rather than being merely a technical tool for the construction of AdS space in string theory.
Moreover, we will see how the above instabilities naturally appear in this context at a perturbative level, through the crucial coupling between the closed and open string sectors underlying a given
AdS stringy construction. 

\subsection*{The brane and the flux pictures}

If we focus on AdS vacua obtained from string compactifications, there are two different approaches that one may choose.
The first is to view the vacua as near-horizon (NH) geometries of a given charged membrane solution, \emph{i.e.} the higher-dimensional
 generalization of the Reissner-Nordstr\"om (RN) black hole in 4D. Some well-known examples in the maximally and half-maximally supersymmetric cases
are collected in table~\ref{branes}.
\begin{table}[h!]
\begin{center}
\scalebox{1}[1]{
\begin{tabular}{| c | c | c |}
\hline
Theory & Brane Type: & Near-Horizon Geometry \\[1mm]
\hline \hline
M-theory & M2 & $\textrm{AdS}_{4}\times S^{7}$ \\
\hline
Type IIB & D3 & $\textrm{AdS}_{5}\times S^{5}$ \\
\hline
M-theory & M5 & $\textrm{AdS}_{7}\times S^{4}$ \\
\hline \hline
Type IIA & D2/D6 & $\textrm{AdS}_{4}\times \tilde{S}^{6}$ \\
\hline
Massive Type IIA & D4/D8 & $\textrm{AdS}_{6}\times \tilde{S}^{4}$ \\
\hline
Type IIB & D3/W & $\textrm{AdS}_{3}\times \tilde{S}^{7}$ \\
\hline
M-theory & M5/KK6 & $\textrm{AdS}_{7}\times \tilde{S}^{4}/\mathbb{Z}_{N}$ \\
\hline
\end{tabular}
}
\end{center}
\caption{{\it Examples of supersymmetric brane systems giving rise to AdS vacua preserving 32 (upper half) and 16 (lower half) supercharges.
In the latter case, the internal space $\tilde{S}^{d}$ denotes a squashed sphere. 
See also \protect\cite{Cvetic:2000cj} for a more complete classification. } 
\label{branes}}
\end{table}
Further possibilities with lower amount of supersymmetry were studied in \cite{Kounnas:2007dd}. An analogous construction is in principle possible in a
non-supersymmetric set-up as well, though it involves further technical complications. 
We will refer to this approach as adopting the \emph{brane picture}.
In this context the focus is on the branes that source the fluxes supporting the AdS vacuum, which emerges when taking the NH limit. Sometimes,
 as is well explained in \cite{Kounnas:2007dd}, such brane backgrounds need special spacetime-filling branes due to charge conservation issues.
However, it remains obscure how these branes will backreact on the geometry sourced by the rest of the brane system and we currently have no
supergravity description thereof.   

The second approach to the search for AdS vacua is that of flux compactifications, where one casts explicit truncation Ans\"atze on the 10D/11D
fields and tries to stabilize all the modes corresponding to geometrical fluctuations of the specific background by means of fluxes threading internal
space. Such fluxes induce a potential for the aforementioned modes and in some cases lead to full moduli stabilization (see \emph{e.g.} 
\cite{DeWolfe:2005uu}). In this picture, fluxes are assumed to be ``God-given'', in the sense that the branes which are supposed to dynamically  generate them do not play any role. In some specific flux backgrounds, non-trivial flux tadpoles (\emph{i.e.} certain quadratic combinations of 
fluxes) are generated and need to be cancelled for consistency by explicitly adding spacetime-filling branes to take care of charge conservation. 
We will refer to this approach as adopting the \emph{flux picture}.
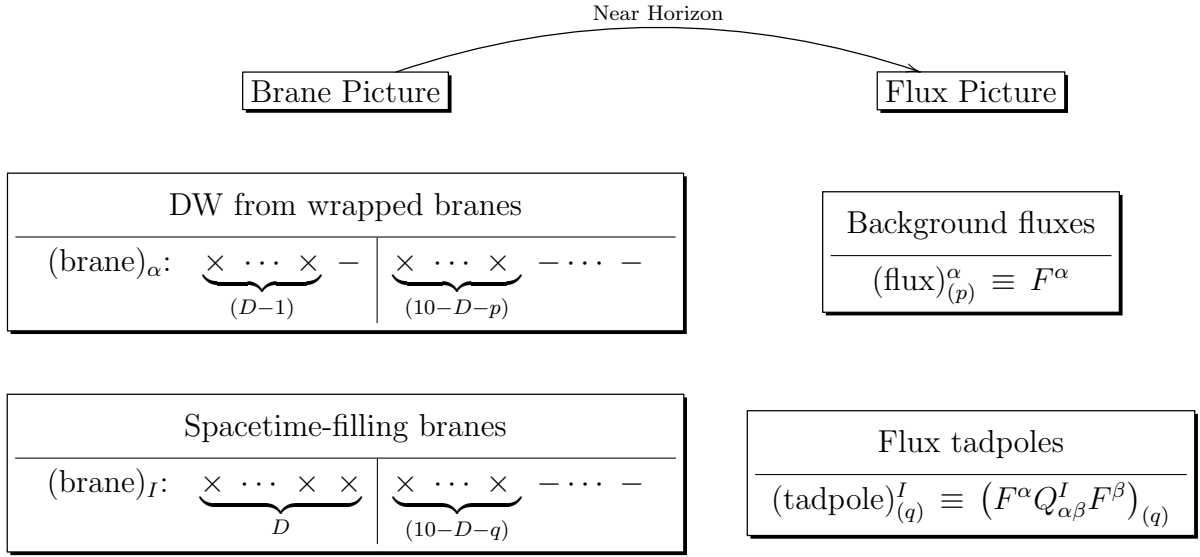
\begin{figure}[h!]
\begin{center}
\scalebox{1}[1]{\xymatrix{ *+[F-,]{\textrm{Brane Picture}} \ar@/^2pc/[r]^{\textrm{Near Horizon}} & *+[F-,]{\textrm{Flux Picture}} \\
*+[F-,]{\begin{tabular}{c}
DW from wrapped branes \\[1mm]
\hline
\begin{tabular}{ c  c | c }
(brane)$_{\alpha}$: & $\underbrace{\times \ \cdots \ \times}_{(D-1)} \ - $  & $\underbrace{\times \ \cdots \ \times \ }_{(10-D-p)}\ - \cdots \ - $
\end{tabular} \end{tabular}} & *+[F-,]{\begin{tabular}{ c }
Background fluxes \\[1mm]
\hline
$(\textrm{flux})^{\alpha}_{(p)}\,\equiv\,F^{\alpha}$
\end{tabular}} \\
*+[F-,]{\begin{tabular}{ c }
Spacetime-filling branes \\[1mm]
\hline
\begin{tabular}{ c  c | c }
(brane)$_{I}$: & $\underbrace{\times \ \cdots \ \times \ \times}_{D}$ & $\underbrace{\times \ \cdots \ \times \ }_{(10-D-q)}\ - \cdots \ - $
\end{tabular} \end{tabular}} & *+[F-,]{\begin{tabular}{ c }
Flux tadpoles \\[1mm]
\hline
$(\textrm{tadpole})^{I}_{(q)}\,\equiv\,\left(F^{\alpha}Q_{\alpha\beta}^{I}F^{\beta}\right)_{(q)}$ \end{tabular}} \\
 }}
\end{center}
\caption{{\it In the brane picture, a given brane system that looks like a DW within the $D$-dimensional theory, may give rise to an AdS$_D$ vacuum in
its near-horizon limit. Furthermore, as discussed in \protect\cite{Kounnas:2007dd}, extra spacetime-filling objects could be required for consistency.
In the flux picture, such an AdS vacuum can be obtained from a compactification supported by $p$-form fluxes wrapping cycles of the internal
manifold. Each of these fluxes is sourced by a corresponding extended object in the brane picture. Possible flux tadpoles are in correspondence with spacetime-filling sources.}\label{fig:brane/flux}} \end{figure}

We conclude that the brane \& the flux pictures provide two dual (hence partial) descriptions of an AdS vacuum. 
The key points of this correspondence are depicted in figure~\ref{fig:brane/flux}. Note that in both descriptions one has access to the same AdS vacuum
though it turns out to be extremely complicated to study the backreaction of the spacetime-filling branes on the background AdS geometry. 
Here we argue that this phenomenon, which may not be neglected in a consistent quantum gravity treatment, is the crucial origin of the universal  instabilities of non-supersymmetric AdS vacua discussed in \cite{Ooguri:2016pdq}.

\subsection*{The brane picture and universal instabilities}
\begin{figure}[h!]
\begin{center}
\scalebox{0.7}[0.7]{\begin{tikzpicture}
\node (I)    at ( 0,0)   {U};
\node (II)    at ( 8,0)   {U$^\prime$};
\node (III)    at ( 4,2)   {W};
\node (A)    at ( -2,2)   {};
\node (B)    at ( 2,2)   {};
\node (C)    at ( 6,2)   {};
\node (D)    at ( 10,2)   {};
\node (E1)    at ( -4,3)   {$\cdots\cdots$};
\node (E2)    at ( 12,3)   {$\cdots\cdots$};
\node (F1)    at ( 0,5)   {};
\node (F2)    at ( 8,5)   {};

\path 
   (I) +(90:4)  coordinate (Itop)
       +(-90:4) coordinate (Ibot)
       +(180:4) coordinate (Ileft)
       +(0:4)   coordinate (Iright)
       ;
\draw  (Ileft) -- 
node[below left, sloped] {$r=Q$}
(Itop) -- 
node[above right, sloped] {$r=Q$}
(Iright) -- 
(Ibot) -- 
(Ileft) -- cycle;

\path 
   (II) +(90:4)  coordinate (IItop)
       +(-90:4) coordinate (IIbot)
       +(180:4) coordinate (IIleft)
       +(0:4)   coordinate (IIright)
       ;
\draw  (IIleft) -- 
node[below right, sloped] {$r=Q$}
(IItop) -- 
node[midway, above, sloped] {$r=Q$}
(IIright) -- 
(IIbot) -- 
(IIleft) -- cycle;

\draw[snake] (Itop) -- (IItop)
      node[midway, above, sloped] {$r=0$};

\draw[thick,dashed] 
    (A) to[out=-45, in=-135, looseness=1.0]  (B);
\draw[thick,dashed] 
    (B) to[out=45, in=135, looseness=1.0] node[midway, above] {$\gamma$} (C);
\draw[->,thick,dashed]
    (C) to[out=-45, in=-135, looseness=1.0]  (D);

\draw[->] (F1) -- (F2)
      node[midway, above, sloped] {time};

\end{tikzpicture}}
\caption{{\it The causal structure of an extremal RN black hole. The oscillatory motion of a free-falling observer around the horizon (dotted timelike curve $\gamma$) necessarily would connect the two 
universes U \& U$^\prime$ through the wormhole W and this structure repeats itself infinitely.}}
\label{fig:Warmhole}
\end{center}
\end{figure}
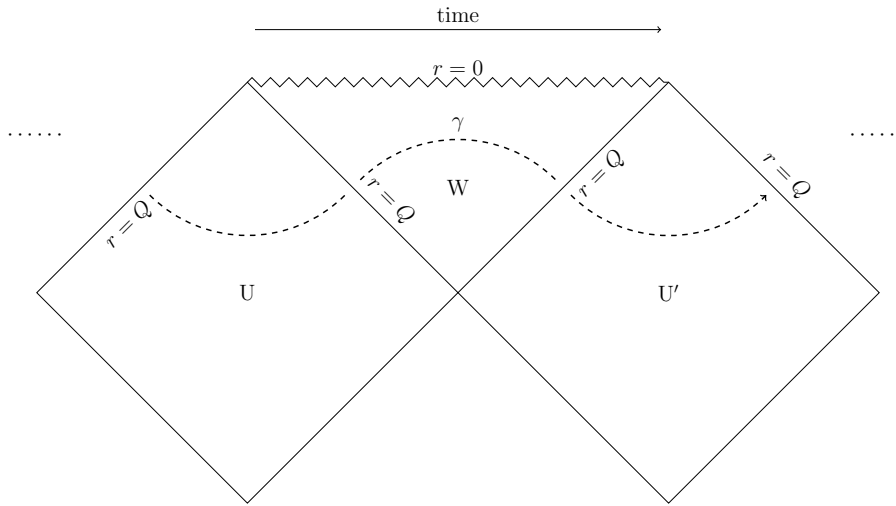

Let us now consider the extremal RN black hole solution in the 4D Einstein-Maxwell theory. The metric and the gauge field read
\begin{equation}
\begin{array}{lclc}
ds_{4}^{2} & = & - \left(1-\frac{Q}{r}\right)^{2}\,dt^{2} \, + \, \frac{dr^{2}}{\left(1-\frac{Q}{r}\right)^{2}} \, + \, r^{2}\,d\Omega_{(2)}^{2} & , 
\\[2mm]
A_{(1)} & = & \frac{Q}{r} \, dt & .
\end{array}
\end{equation}
In the NH limit (\emph{i.e.} when taking $r\rightarrow Q$), the above metric behaves as $\textrm{AdS}_{2}\times S^{2}$. This tells us that, by adopting
the brane picture, the RN solution can be seen as a 0-brane solution whose NH limit gives rise to an $\textrm{AdS}_{2}$ vacuum of the theory supported by an $\tilde{F}_{(2)}$ flux wrapping $S^{2}$, which
is associated with $*_{4}(dA_{(1)})$. Solving the geodesic equations we find a motion described by 
\be
y(\tau)\,=\,\alpha\,Q\,\sin\left(\frac{\tau}{Q}\right) \ ,
\ee
where $\tau$ denotes proper time for a free-falling observer, $y\,\equiv\,(r-Q)$, and $\alpha$ is a constant. 
The condition $y\,\overset{!}{\ll}\,Q$, \emph{i.e.} $\alpha\,\overset{!}{\ll}\,1$ makes sure we stay in AdS. 
This motion will take us through the wormholes into other universes as illustrated in figure~\ref{fig:Warmhole}. Note that an observer within AdS will only stay in a given universe for at most half a period of such oscillatory motion, and a process of duration $\Delta t$
in the brane picture will blueshift to $\Delta\tau\,\approx\,\alpha\,\Delta t\,\rightarrow\,0$ near the horizon. 

\begin{figure}[h!]
\begin{center}
\scalebox{0.7}[0.7]{\begin{tikzpicture}
\node (I)    at ( 4,0)   {I};
\node (II)    at ( 1.7,4.2)   {II};
\node (IIp)    at ( 0,0)   {};
\node (IIs)    at ( 0,3.5)   {};

\path 
   (IIp) +(90:8)  coordinate (IItop)
         +(45:5.66)   coordinate (IIcen)
         +(0:0) coordinate (IIbot);

\draw  (IIbot) -- (IIcen) -- (IItop); 
\draw[snake] (IIbot) -- (IItop)
      node[midway, above, sloped] {$r=0$};

\path 
   (I) +(90:4)  coordinate[label=90:$i_+$] (Itop)
       +(-90:4) coordinate (Ibot)
       +(180:4) coordinate[label=180:$i_\infty$] (Ileft)
       +(0:4)   coordinate[label=0:$i_0$] (Iright)
       ;
\draw  (Ileft) -- 
node[midway, below, sloped] {$r=Q$}
(Itop) -- 
node[midway, above, sloped] {$\mathcal{I}^+$}
(Iright) -- 
(Ibot) -- 
node[midway, below, sloped] {$r=Q$}
(Ileft) -- cycle;

\draw[->] 
    (Iright) to[out=165, in=10, looseness=1.5] node[midway, below] {$\Sigma_0$} (IIs);
\end{tikzpicture}}
\caption{{\it The example of an initial data surface $\Sigma_0$ for a charged scalar field in the extremal RN geometry. It may be noted that these surfaces are universally incomplete 
\protect\cite{Kunduri:2013ana}.}}
\label{fig:Penrose}
\end{center}
\end{figure}
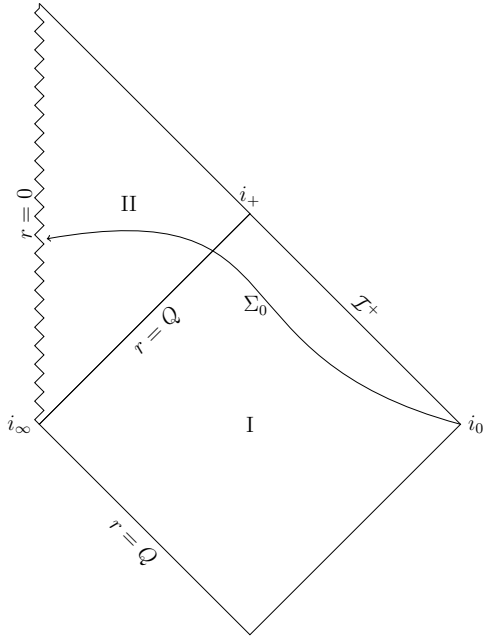

In refs~\cite{Aretakis:2011ha,Aretakis:2011hc,Lucietti:2012xr} the NH $\textrm{AdS}_{2}\times S^{2}$ geometry was shown to be unstable w.r.t. perturbations associated with a charged scalar field (both
massless \& massive). The analysis can be done perturbatively at a classical level and the resulting instability agrees with the intuition that such a background should develop perturbations that create 
a trapped surface, which, according to the singularity theorems, results in a geodesically incomplete spacetime. This is well illustrated in figure~\ref{fig:Penrose}, where we show that the initial data
surfaces for scalar fields in the extremal RN geometry are not complete. 

%{\bf WORMHOLES AND OTHER UNIVERSES SHOULD BE ADDED IN PICTURE}
%

It is reasonable to expect that any consistent theory of quantum gravity would save us from the above disastrous conclusion that the system will develop a naked singularity. To understand how this puzzle can be resolved,
we may view the above instability from the perspective of the KK reduction of 4D gravity on $S^{2}$ \cite{Maldacena:1998uz}, where this arises from taking the backreaction of matter into account.  
Following the argument in \cite{Durkee:2010ea}, this instability of the NH geometry implies an instability of the extremal RN black hole through Schwinger pair production.
This way of reasoning is fully consistent with \cite{Ooguri:2016pdq}, since these particle/anti-particle pairs must be \emph{microscopic} (in the sense that they violate the BPS bound). 
Moreover, they represent a channel of non-perturbative instability for our ``0-brane solution'', and this in turn leads to a perturbative instability of the corresponding AdS$_2$ vacuum due to 
an infinite redshift effect at the horizon that effectively shrinks its lifetime to zero. Interestingly, it was observed in \cite{Lucietti:2012xr} that the instability of the extremal RN black hole may be captured within the NH limit thanks to the crucial coupling between
gravitational and electromagnetic perturbations. 

Motivated by this 4D example, we will now discuss the origin of this type of instability in the case of stringy brane constructions giving rise to 
higher-dimensional AdS space. Going back to the generic situation in the brane picture, all the branes in the upper line of figure~\ref{fig:brane/flux} (\emph{i.e.} those ones which are not spacetime-filling!) do give rise to a 
supersymmetric brane configuration. Now, depending on the different cases, the addition of spacetime-filling branes (lower line in figure~\ref{fig:brane/flux}) might be required or not for consistency.
In this way, one obtains a brane construction giving rise to a supersymmetric AdS vacuum in the NH limit. What needs to be discussed now is the supersymmetry breaking mechanism.
 
In ref.~\cite{Argurio:2007qk}, a concrete mechanism for dynamical supersymmetry breaking was proposed when considering a special type IIA setting involving NS5-branes and D4-branes. 
The construction is based on the possibility of performing a Higgs branch movement consisting in tilting one stack of NS5-branes w.r.t. a neighboring one. 
Subsequently, one acts with a Seiberg duality \cite{Seiberg:1997vw} that exchanges two consecutive NS5 stacks that are mutually tilted. This results in a net creation of $\overline{\rm D4}$-branes and
describes a supersymmetry breaking mechanism, which is interpreted in the dual field theory as turning on mesonic vev's.

The reason why this new non-supersymmetric brane configuration is metastable is due to the fact that the created $\overline{\rm D4}$'s are separated in space from the D4 stack, and brane/anti-brane
annihilation will only occur at a non-perturbative level through nucleation of microscopic NS5 bubbles. However, when reaching the conformal fixed point by letting the NS5 stacks collide, the
D4's \& $\overline{\rm D4}$'s will now sit on top of each other and hence will annihilate perturbatively. It is important to note that such a perturbative decay channel crucially needs the
exchange of open string degrees of freedom which become tachyonic. If one restricts to the pure closed string sector, there will just be a static DW separating the non-supersymmetric vacuum and 
the supersymmetric one obtained after brane/anti-brane annihilation. 

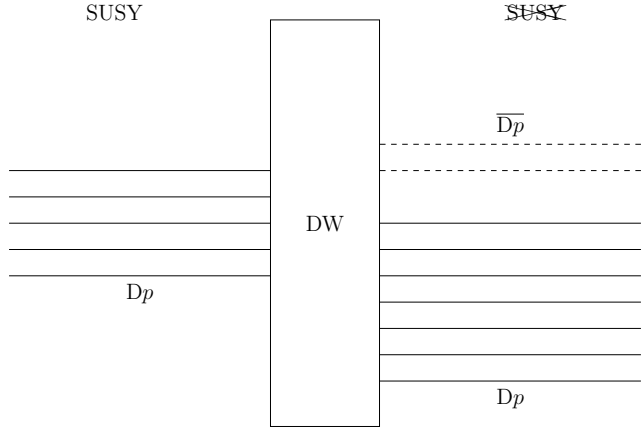
\begin{figure}[h!]
\begin{center}
\scalebox{0.7}[0.7]{\begin{tikzpicture}
\node (I)    at (0,0)   {DW};
\node (II)    at (-4,4)   {SUSY};
\node (III)    at (4,4)   {$\xcancel{\rm SUSY}$};

\path 
   (I) +( 75:4) coordinate (IRT)
       +(105:4) coordinate (ILT)
       +(255:4) coordinate (ILB)
       +(-75:4) coordinate (IRB)
       ;
\draw  (IRT) -- (ILT) -- (ILB) -- (IRB) -- cycle;

\draw  (-6,1) -- (-1.035,1);
\draw  (-6,0.5) -- (-1.035,0.5);
\draw  (-6,0) -- (-1.035,0);
\draw  (-6,-0.5) -- (-1.035,-0.5);
\draw  (-6,-1) -- node[midway, below] {D$p$} (-1.035,-1);

\draw[dashed]  (6,1.5) -- node[midway, above] {$\overline{{\rm D}p}$} (1.035,1.5);
\draw[dashed]  (6,1) -- (1.035,1);

\draw  (6,0) -- (1.035,0);
\draw  (6,-0.5) -- (1.035,-0.5);
\draw  (6,-1) -- (1.035,-1);
\draw  (6,-1.5) -- (1.035,-1.5);
\draw  (6,-2) -- (1.035,-2);
\draw  (6,-2.5) -- (1.035,-2.5);
\draw  (6,-3) -- node[midway, below] {D$p$} (1.035,-3);

\end{tikzpicture}}
\caption{{\it An artist's impression of the difference between supersymmetric (left) and non-supersymmetric (right) AdS vacua as they are constructed in the brane picture. The thick static DW separating them
contains all the information concerning the branes which are not spacetime-filling. Note that the brane/anti-brane annihilation process may perturbatively occur on the right side of this picture only
upon including the open string degrees of freedom living on the $p$-branes.}}
\label{fig:SUSY_breaking}
\end{center}
\end{figure}

Exporting this intuition to a generic setting, we propose that the brane system realizing a certain class of flux backgrounds is universally independent of supersymmetry.
In such a case, the difference between the brane construction underlying a supersymmetric and a non-supersymmetric AdS vacuum would only lie in the type of spacetime-filling sources which one adds to 
cancel the tadpoles. Sources carrying pure brane-charge correspond to a supersymmetric realization, while a mixture of sources carrying brane \& anti-brane charge corresponds to a non-supersymmetric
one. A conceptual picture of this idea is shown in figure~\ref{fig:SUSY_breaking}.

\subsection*{Proposing an effective description}

In the last subsection we have shown how the difference between supersymmetric and non-supersymmetric brane systems giving rise to AdS vacua in the NH limit
is related to the presence, in this latter case, of brane/anti-brane pairs probing the geometry sourced by the rest of the brane system. All of this, 
effectively looks like a non-BPS DW from the viewpoint of an effective $D$-dimensional gravity description. 

In the limit where one neglects the interaction
between spacetime-filling branes and the background, the relevant effective description is exactly given by a $D$-dimensional gauged supergravity theory
where supersymmetric and non-supersymmetric AdS vacua are (possibly) separated by a static non-BPS DW. Within such a truncation, which corresponds to a restriction to the closed string excitations, one is led to the conclusion that no gravitational tunneling will occur and hence that the non-supersymmetric
vacuum is non-perturbatively stable (see \cite{Danielsson:2016rmq} for the details of the argument). 

Adopting the completely opposite perspective of analyzing the dynamical evolution of a stack of $N$ spacetime-filling branes, very important results may be
obtained in the large $N$ limit, \emph{i.e.} when the influence of the above geometrical background on the open string degrees of freedom can be safely neglected. As an example of this, recently in \cite{Bena:2016oqr} this avenue has been pursued in order to analyze the polarization process of spacetime-filling branes.

Once again, both descriptions are interesting and relevant in a particular regime. However, the instabilities that we have discussed are crucial
consequences of the coupling between the open and closed string sectors. We remind the reader that even in cases of AdS solutions obtained without need for
local sources (see \emph{e.g.} \cite{Dibitetto:2011gm}), it is inconsistent to neglect the effect of the open string sector when embedding the effective description at hand into its UV completion \cite{Green:2007zzb}.

As already argued, these instabilities arising within the brane picture from the nucleation of microscopic brane bubbles, do assume a perturbative
character in the NH limit. We propose that such processes could be captured by an effective gauged supergravity model in $D$ dimensions, where the 
universal sector describing closed string excitations is coupled to $N$ extra vector multiplets describing the dynamical degrees of freedom localized on
the spacetime-filling branes. 

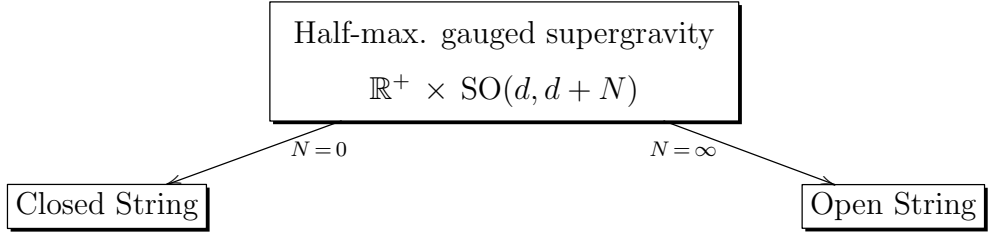
\begin{figure}
\begin{center}
\scalebox{1}[1]{\xymatrix{ &  *+[F-,]{\begin{tabular}{c}
Half-max. gauged supergravity \\[1mm]
$\mathbb{R}^+\,\times\,\textrm{SO}(d,d+N)$
\end{tabular}}\ar[dl]^{N\,=\,0}\ar[dr]_{N\,=\,\infty} &  \\
 *+[F-,]{\textrm{Closed String}} & & *+[F-,]{\textrm{Open String}} \\
 }}
\end{center}
\caption{{\it The effective gauged supergravity theory describing the coupling between the closed and open string sector of a given compactification. In the $N\,=\,0$ limit (left) AdS flux vacua are generically separated by static DW's and hence do not show instabilities. In the $N\,=\,\infty$ limit, one can adopt the description of spacetime-filling branes as probes. The actual calculation that shows the presence of a tachyon at a perturbative level is the one at finite $N$ in this picture, \emph{i.e.} where no decoupling limit is accessible.}\label{fig:effective_model}} \end{figure}

As an example, let us consider a string compactification on a $\mathbb{T}^{10-D}$ with fluxes supported by a single type of spacetime-filling sources. 
In this case the effective theory describing this would be a half-maximal gauged supergravity in $D$ dimensions coupled to $d+N$ vector multiplets,
where $N$ is the number of branes composing the stack. The corresponding situation can then be represented as in figure~\ref{fig:effective_model}.
We plan to discuss the validity of this approach in a future work \cite{WIP} within a concrete example where calculations at finite $N$ can be performed in
a precise way.

\section{Discussion}

In this paper we have investigated the issue of stability of non-supersymmetric AdS vacua admitting a consistent UV completion within string theory.
Our approach was inspired by a stronger version of the WGC formulated in \cite{Ooguri:2016pdq}, where it was further used to conclude that any non-supersymmetric membrane configuration in 
a qunatum theory of gravity will eventually decay through emission of microscopic, charged particles. The ultimate consequence of their proposal is that all non-supersymmetric AdS vacua that may be constructed in this way
are unstable.

In this paper we have taken a further step in this direction, and spelled out a correspondence between the brane and the flux picture as two \emph{equivalent} ways of constructing AdS vacua in string theory.
In the former picture, they arise as NH geometries of membrane intersections, whereas in the latter one they are obtained as vacua supported by fluxes threading cycles of a given compact manifold.
Our take-home message would be that viewing AdS vacua as the NH limit of membranes is not merely a technical trick, but rather a physical statement and, as such, it may be used to analyze the 
stability of all AdS vacua.

We started out by reviewing the instability arguments for the 4D extremal RN black hole, which were already used in the GR literature to conclude the instability of the $\textrm{AdS}_{2}\times S^{2}$ NH geometry. 
It may be worth highlighting that, while the $\textrm{AdS}_{2}$ instability at the horizon has a perturbative nature, the instability of the black hole geometry in itself is non-perturbative and possibly
related to Schwinger pair production, as proposed long ago in \cite{Maldacena:1998uz}.
Note that all of these arguments crucially make use of the presence of charged matter coupled to geometry.  

We then moved on to higher-dimensional generalizations of the RN black hole construction underlying the $\textrm{AdS}_{2}$ vacuum of the Einstein-Maxwell theory. 
A collection of interesting examples of AdS string vacua constructed from branes can be found in \cite{Kounnas:2007dd}. The important feature of these models is that all branes that source fluxes in
internal space reduce to a BPS DW giving rise to the AdS vacuum in the NH limit, while extra spacetime-filling branes generically need to be added for tadpole cancellation to work.
We do not know how to properly take the backreaction of these branes on the background into account.

Nevertheless, we have followed the lines of \cite{Argurio:2007qk} to motivate a universal mechanism for supersymmetry breaking. Based on this, the only difference between a supersymmetric and a 
non-supersymmetric configuration in the brane picture consists in realizing the tadpole produced by the non-spacetime-filling branes in terms of pure spacetime-filling branes rather than branes \& 
anti-branes at the same time. In such a context, the universal channel of perturbative instability is expected within the open string sector and corresponds to the tachyon responsible for brane/anti-brane
annihilation.

Note that, in general, this tachyon lies outside of the truncation to the pure closed string sector, which was recently discussed in \cite{Danielsson:2016rmq}. Within such an effective description, specific 
examples of non-supersymmetric AdS vacua were discussed, which turn out to be even non-perturbatively stable.  
However, one needs to be very careful when embedding a concrete model within a proper UV complete description. This kind of truncation is actually 
inconsistent from the viewpoint of string theory \cite{Green:2007zzb}, since extended objects with open strings attached do belong to the spectrum of the theory. 
 
Our conclusion is, therefore, that all non-supersymmetric AdS vacua suffer from this universal instability channel which relies on the crucial interaction between the closed and the open string sectors
that is necessarily present in a quantum gravity regime. Due to this intrinsically quantum feature, it is generically very hard to compute these effects. Nevertheless, we propose a rather simple
setting where the main features of this phenomenon can be captured within a classical though lower-dimensional theory retaining some of the stringy properties of the model. 
We hope to be able to test this proposal in concrete examples in the future \cite{WIP}.

In line with \cite{Ooguri:2016pdq}, we view the universal instabilities discussed here as an inevitable consequence of the WGC, but we expect none of this to occur in a supersymmetric case. 
However, it would be nice to further understand the relation to the recent work done in \cite{Freivogel:2016qwc}, where even possible instabilities of supersymmetric AdS solutions are discussed.
It still remains to clarify what all of this could mean for non-supersymmetric holographic constructions 
(see \emph{e.g.} \cite{Argurio:2007qk,Apruzzi:2016rny}). Interestingly, the holographic paradigm per se relies on a decoupling limit between open \& closed strings where our results do not apply.
This suggests that a full quantum gravity description might be irrelevant for holography, which then would be meaningful only in the decoupling limit. Alternatively, one could put one's hopes to  $\mathcal{N}=8$ SUGR. If this theory
is UV-complete all on its own, independent of string theory, it would be a candidate framework for non-supersymmetric holography without reservations.

The next very important step is to establish what the consequences are for constructions of dS. In particular, we believe there is a connection with the instabilities discussed in e.g. \cite{Blaback:2012nf} (for a recent paper with references, see  \cite{Danielsson:2016ci}), which are of the same type as those relevant to the WGC. We hope to return to this in the near future. \footnote{We thank Thomas Vant Riet for interesting discussions about this.}

\section*{Acknowledgments}

We would like to thank Cliff Burgess, Matthew Kleban, Ben Freivogel, Sergio Vargas, and Thomas Van Riet for very interesting and stimulating discussions.
The work of the authors was supported by the Swedish Research Council (VR).

%%%%%%%%%%%%%%%%%%%%%%%%%%%%%%%%%%%%
%
% Bibliography
%
%%%%%%%%%%%%%%%%%%%%%%%%%%%%%%%%%%%%

\small

%\clearpage

\bibliography{references}

\providecommand{\href}[2]{#2}\begingroup\raggedright\begin{thebibliography}{10}

\bibitem{Maldacena:1997re}
J.~M. Maldacena, ``{The Large N limit of superconformal field theories and
  supergravity},'' \href{http://dx.doi.org/10.1023/A:1026654312961}{{\em Int.
  J. Theor. Phys.} {\bf 38} (1999)  1113--1133},
  \href{http://arxiv.org/abs/hep-th/9711200}{{\tt arXiv:hep-th/9711200
  [hep-th]}}.
[Adv. Theor. Math. Phys.2,231(1998)].
%%CITATION = HEP-TH/9711200;%%.

\bibitem{Argurio:2007qk}
R.~Argurio, M.~Bertolini, S.~Franco, and S.~Kachru, ``{Meta-stable vacua and
  D-branes at the conifold},''
  \href{http://dx.doi.org/10.1088/1126-6708/2007/06/017}{{\em JHEP} {\bf 06}
  (2007)  017},
\href{http://arxiv.org/abs/hep-th/0703236}{{\tt arXiv:hep-th/0703236
  [hep-th]}}.
%%CITATION = HEP-TH/0703236;%%.

\bibitem{Ooguri:2016pdq}
H.~Ooguri and C.~Vafa, ``{Non-supersymmetric AdS and the Swampland},''
\href{http://arxiv.org/abs/1610.01533}{{\tt arXiv:1610.01533 [hep-th]}}.
%%CITATION = ARXIV:1610.01533;%%.

\bibitem{Freivogel:2016qwc}
B.~Freivogel and M.~Kleban, ``{Vacua Morghulis},''
\href{http://arxiv.org/abs/1610.04564}{{\tt arXiv:1610.04564 [hep-th]}}.
%%CITATION = ARXIV:1610.04564;%%.

\bibitem{Kounnas:2007dd}
C.~Kounnas, D.~Lust, P.~M. Petropoulos, and D.~Tsimpis, ``{AdS4 flux vacua in
  type II superstrings and their domain-wall solutions},''
  \href{http://dx.doi.org/10.1088/1126-6708/2007/09/051}{{\em JHEP} {\bf 09}
  (2007)  051},
\href{http://arxiv.org/abs/0707.4270}{{\tt arXiv:0707.4270 [hep-th]}}.
%%CITATION = ARXIV:0707.4270;%%.

\bibitem{Lucietti:2012xr}
J.~Lucietti, K.~Murata, H.~S. Reall, and N.~Tanahashi, ``{On the horizon
  instability of an extreme Reissner-Nordstr\'om black hole},''
  \href{http://dx.doi.org/10.1007/JHEP03(2013)035}{{\em JHEP} {\bf 03} (2013)
  035},
\href{http://arxiv.org/abs/1212.2557}{{\tt arXiv:1212.2557 [gr-qc]}}.
%%CITATION = ARXIV:1212.2557;%%.

\bibitem{Aretakis:2011ha}
S.~Aretakis, ``{Stability and Instability of Extreme Reissner-Nordstr\'om Black
  Hole Spacetimes for Linear Scalar Perturbations I},''
  \href{http://dx.doi.org/10.1007/s00220-011-1254-5}{{\em Commun. Math. Phys.}
  {\bf 307} (2011)  17--63},
\href{http://arxiv.org/abs/1110.2007}{{\tt arXiv:1110.2007 [gr-qc]}}.
%%CITATION = ARXIV:1110.2007;%%.

\bibitem{Aretakis:2011hc}
S.~Aretakis, ``{Stability and Instability of Extreme Reissner-Nordstrom Black
  Hole Spacetimes for Linear Scalar Perturbations II},''
  \href{http://dx.doi.org/10.1007/s00023-011-0110-7}{{\em Annales Henri
  Poincare} {\bf 12} (2011)  1491--1538},
\href{http://arxiv.org/abs/1110.2009}{{\tt arXiv:1110.2009 [gr-qc]}}.
%%CITATION = ARXIV:1110.2009;%%.

\bibitem{Danielsson:2016rmq}
U.~H. Danielsson, G.~Dibitetto, and S.~C. Vargas, ``{Universal isolation in the
  AdS landscape},''
\href{http://arxiv.org/abs/1605.09289}{{\tt arXiv:1605.09289 [hep-th]}}.
%%CITATION = ARXIV:1605.09289;%%.

\bibitem{Banks:2002nm}
T.~Banks, ``{Heretics of the false vacuum: Gravitational effects on and of
  vacuum decay. 2.},''
\href{http://arxiv.org/abs/hep-th/0211160}{{\tt arXiv:hep-th/0211160
  [hep-th]}}.
%%CITATION = HEP-TH/0211160;%%.

\bibitem{Green:2007zzb}
M.~B. Green, H.~Ooguri, and J.~H. Schwarz, ``{Nondecoupling of Maximal
  Supergravity from the Superstring},''
  \href{http://dx.doi.org/10.1103/PhysRevLett.99.041601}{{\em Phys. Rev. Lett.}
  {\bf 99} (2007)  041601},
\href{http://arxiv.org/abs/0704.0777}{{\tt arXiv:0704.0777 [hep-th]}}.
%%CITATION = ARXIV:0704.0777;%%.

\bibitem{Cvetic:2000cj}
M.~Cvetic, H.~Lu, C.~N. Pope, and J.~F. Vazquez-Poritz, ``{AdS in warped
  space-times},'' \href{http://dx.doi.org/10.1103/PhysRevD.62.122003}{{\em
  Phys. Rev.} {\bf D62} (2000)  122003},
\href{http://arxiv.org/abs/hep-th/0005246}{{\tt arXiv:hep-th/0005246
  [hep-th]}}.
%%CITATION = HEP-TH/0005246;%%.

\bibitem{DeWolfe:2005uu}
O.~DeWolfe, A.~Giryavets, S.~Kachru, and W.~Taylor, ``{Type IIA moduli
  stabilization},'' \href{http://dx.doi.org/10.1088/1126-6708/2005/07/066}{{\em
  JHEP} {\bf 07} (2005)  066},
\href{http://arxiv.org/abs/hep-th/0505160}{{\tt arXiv:hep-th/0505160
  [hep-th]}}.
%%CITATION = HEP-TH/0505160;%%.

\bibitem{Kunduri:2013ana}
H.~K. Kunduri and J.~Lucietti, ``{Classification of near-horizon geometries of
  extremal black holes},'' \href{http://dx.doi.org/10.12942/lrr-2013-8}{{\em
  Living Rev. Rel.} {\bf 16} (2013)  8},
\href{http://arxiv.org/abs/1306.2517}{{\tt arXiv:1306.2517 [hep-th]}}.
%%CITATION = ARXIV:1306.2517;%%.

\bibitem{Maldacena:1998uz}
J.~M. Maldacena, J.~Michelson, and A.~Strominger, ``{Anti-de Sitter
  fragmentation},'' \href{http://dx.doi.org/10.1088/1126-6708/1999/02/011}{{\em
  JHEP} {\bf 02} (1999)  011},
\href{http://arxiv.org/abs/hep-th/9812073}{{\tt arXiv:hep-th/9812073
  [hep-th]}}.
%%CITATION = HEP-TH/9812073;%%.

\bibitem{Durkee:2010ea}
M.~Durkee and H.~S. Reall, ``{Perturbations of near-horizon geometries and
  instabilities of Myers-Perry black holes},''
  \href{http://dx.doi.org/10.1103/PhysRevD.83.104044}{{\em Phys. Rev.} {\bf
  D83} (2011)  104044},
\href{http://arxiv.org/abs/1012.4805}{{\tt arXiv:1012.4805 [hep-th]}}.
%%CITATION = ARXIV:1012.4805;%%.

\bibitem{Seiberg:1997vw}
N.~Seiberg, ``{The Power of duality: Exact results in 4-D SUSY field theory},''
  \href{http://dx.doi.org/10.1142/S0217751X01005705, 10.1142/S0217751X97002772,
  10.1143/PTPS.123.337}{{\em Int. J. Mod. Phys.} {\bf A16} (2001)  4365--4376},
  \href{http://arxiv.org/abs/hep-th/9506077}{{\tt arXiv:hep-th/9506077
  [hep-th]}}.
[Prog. Theor. Phys. Suppl.123,337(1996)].
%%CITATION = HEP-TH/9506077;%%.

\bibitem{Bena:2016oqr}
I.~Bena, J.~Bl{\aa}b{\"a}ck, R.~Minasian, and R.~Savelli, ``{There and back
  again: A T-brane's tale},''
\href{http://arxiv.org/abs/1608.01221}{{\tt arXiv:1608.01221 [hep-th]}}.
%%CITATION = ARXIV:1608.01221;%%.

\bibitem{Dibitetto:2011gm}
G.~Dibitetto, A.~Guarino, and D.~Roest, ``{Charting the landscape of N=4 flux
  compactifications},'' \href{http://dx.doi.org/10.1007/JHEP03(2011)137}{{\em
  JHEP} {\bf 03} (2011)  137},
\href{http://arxiv.org/abs/1102.0239}{{\tt arXiv:1102.0239 [hep-th]}}.
%%CITATION = ARXIV:1102.0239;%%.

\bibitem{WIP}
U.~Danielsson and G.~Dibitetto, ``{Work in progress},''.

\bibitem{Apruzzi:2016rny}
F.~Apruzzi, G.~Dibitetto, and L.~Tizzano, ``{A new 6d fixed point from
  holography},''
\href{http://arxiv.org/abs/1603.06576}{{\tt arXiv:1603.06576 [hep-th]}}.
%%CITATION = ARXIV:1603.06576;%%.

\bibitem{Blaback:2012nf}
J.~Blaback, U.~H. Danielsson, and T.~Van~Riet, ``{Resolving anti-brane
  singularities through time-dependence},''
  \href{http://dx.doi.org/10.1007/JHEP02(2013)061}{{\em JHEP} {\bf 02} (2013)
  061},
\href{http://arxiv.org/abs/1202.1132}{{\tt arXiv:1202.1132 [hep-th]}}.
%%CITATION = ARXIV:1202.1132;%%.

\bibitem{Danielsson:2016ci}
U.~H. Danielsson, F.~F. Gautason, and T.~Van~Riet, ``{Unstoppable brane-flux
  decay of $\overline{\text{D6}}$ branes},''
\href{http://arxiv.org/abs/1609.06529}{{\tt arXiv:1609.06529 [hep-th]}}.
%%CITATION = ARXIV:1609.06529;%%.

\end{thebibliography}\endgroup
\bibliographystyle{utphys}

\end{document}